\documentclass[11pt]{article}
\usepackage{amssymb}
\usepackage{amsmath}
\usepackage{graphicx}
\usepackage{hep}
\usepackage{sbl_th}
\usepackage{datetime}
\usepackage{amssymb, axodraw}
\usepackage{amsmath}
\usepackage{graphicx}
\usepackage{hep}
\usepackage{color}
\newcommand{\hzero}{\ensuremath{\PHiggslightzero}} 
\newcommand{\Hzero}{\ensuremath{\PHiggsheavyzero}} 
\newcommand{\Azero}{\ensuremath{\PHiggspszero}} 
\newcommand{\Hpm}{\ensuremath{\PHiggspm}} 
\newcommand{\CP}{\ensuremath{\mathcal{C}\mathcal{P}}}

\begin{document}
\phantom{x}
\begin{center}
\vspace{1.3cm}
\begin{center}
\vskip 1.5cm
    {\large \textsc{One-loop Higgs boson production at the Linear Collider within the general two-Higgs-doublet model:
 $\APelectron\Pelectron$ versus $\Pphoton\Pphoton$}}

    \vskip 0.7cm

\textbf{Joan Sol\`a $^{a}$ and David L\'opez-Val$^{b}$}

\vskip 0.7cm

\textit{$^{a}$ High Energy Physics Group, Dept. ECM, and Institut de Ci{\`e}ncies del Cosmos\\
Univ. de Barcelona, Av. Diagonal 647, E-08028 Barcelona, Catalonia, Spain}\\

\vskip5mm

\textit{$^{b}$ Institut f\"ur Theoretische Physik,Universit\"at Heidelberg\\
Philosophenweg 16, D-69120 Heidelberg, Germany}\\

\vskip5mm

E-mails: sola@ecm.ub.es, lopez@thphys.uni-heidelberg.de. \vskip15mm

\end{center}

\begin{quotation}
\noindent {\large\it \underline{Abstract}}.\ \ We present an updated
overview on the phenomenology of one-loop Higgs boson production at
Linear Colliders within the general Two-Higgs-Doublet Model (2HDM).
First we report on the Higgs boson pair production, and associated
Higgs-Z boson production, at $\mathcal{O}(\alpha^3_{ew})$ from
$\APelectron\Pelectron$ collisions. These channels furnish
cross-sections in the range of $10-100\,\femtobarn$ for $\sqrt{s} =
0.5 \,\TeV$ and exhibit potentially large radiative corrections
($|\delta_r|\sim 50\%$), whose origin can be traced back to the
genuine enhancement capabilities of the triple Higgs boson
self-interactions. Next we consider the loop-induced production of a
single Higgs boson from direct $\Pphoton\Pphoton$ scattering. We
single out sizable departures from the expected $\gamma\gamma \to h$
rates in the Standard Model, which are again correlated to trademark
dynamical features of the 2HDM -- namely the balance of the
non-standard Higgs/gauge, Higgs/fermion and Higgs self-interactions
leading to sizable (destructive) interference effects. This pattern
of quantum effects is unmatched in the MSSM, and could hence provide
distinctive footprints of non-supersymmetric Higgs boson physics.
Both calculations are revisited within a common, brought-to-date
framework and include, in particular, the most stringent bounds from
unitarity and flavor physics.
\end{quotation}
\vskip 8mm
\end{center}

\newpage

\section{Introduction and computational framework}

The quest for experimental evidence of the Higgs boson is actively
underway at the Tevatron and the LHC \cite{haberrev}. Nonetheless, a
complete understanding of the Electroweak Symmetry Breaking
conundrum will not only demand to discover the Higgs boson, but also
to precisely measure its mass, quantum numbers and interactions to
the other particles. A Linear Collider (linac, hereafter), such as
the ILC or the CLIC, would be the most natural facility to carry
this endeavor to completion \cite{ILCPhysics} and, perhaps most
significantly, to disentangle the Standard Model (SM) Higgs
mechanism from its many potential extensions.

The Two-Higgs-Doublet Model (2HDM) \cite{2hdmrev} constitutes a
singularly simple, and yet phenomenologically very rich example of
the latter. The absence of tree-level flavor changing neutral
currents determines the Higgs/fermion Yukawa couplings and leads to
the canonical type-I and type-II realizations of the 2HDM
~\cite{2hdmrev}. Moreover, it effectively accounts for the
low-energy Higgs sector of some more fundamental theories, such as
the Minimal Supersymmetric Standard Model (MSSM) \cite{mssm}. The
2HDM can be fully specified in terms of the masses of the physical
Higgs particles; the parameter $\tan \beta$ (the ratio $\langle
H_2^0\rangle/\langle H_1^0\rangle$ of the two VEV's giving masses to
the up- and down-like quarks); the mixing angle $\alpha$ between the
two $\CP$-even states; and, finally, of one genuine Higgs boson
self-coupling, usually denoted as $\lambda_5$. We refer the reader
to Ref.\,\cite{loop1} for full details on the model setup, our
notation, definitions and various constraints.

While in the context of the MSSM we expect a panoply of Yukawa, and
Yukawa-like, couplings of various kinds (including squark
interactions with the Higgs bosons), whose phenomenological
implications have been exploited in the past in a variety of
important processes (see e.g.\,\cite{SUSYenhanced}), in the case of
the general 2HDM we count on alternative mechanisms. Above all, the
Higgs self-interactions are perhaps the very trademark structure of
the 2HDM. Unlike their MSSM counterparts, the triple ($3h$) and
quartic ($4h$) Higgs self-couplings are not restricted by the gauge
symmetry, and so they can be potentially enhanced. In favorable
circumstances, these enhancements can translate into highly
distinctive signatures of a non-standard, non-supersymmetric Higgs
sector. Dedicated literature on the topic includes e.g. the
tree-level studies on triple Higgs boson production,
$\APelectron\Pelectron \to 3h$ \cite{giancarlo}, inclusive
Higgs-pair production through gauge boson fusion,
$\APelectron\Pelectron \to V^*V^* \to 2h+X$\cite{neil}, and the
double Higgs-strahlung channels $\APelectron\Pelectron \to hh Z^0$
\cite{arhrib}. Also significant are the loop-induced single Higgs
production, $\gamma\gamma \to h$ \cite{photon} and the double Higgs
channels $\gamma\gamma \to 2h$ \cite{doublephoton}. They have both
been considered in the framework of a photon-photon collider,
alongside with the complementary radiative decay mode $h \to
\gamma\gamma$ \cite{posch}. Finally, the impact of these $3h$
self-couplings (see e.g. Table II of \cite{loop1}) has been
quantified at the level of radiative corrections through the
detailed one-loop analysis of the pairwise production of both
charged \cite{hw} and neutral \cite{loop1} Higgs boson pairs, as
well as upon the study of the associated Higgs-strahlung channels
$\APelectron\Pelectron \to \hzero Z^0, \Hzero Z^0$ at the quantum
level\,\cite{hw,loop2} \footnote{For related work in the context of
radiative corrections in Higgs production processes, see
e.g.~\cite{mssmloop}. Phenomenological prospects for the LHC have
been addressed in\,\cite{recent2hdm}.}. The effective enhancing
power of the Higgs self-interactions is subdued, in practice, by a
number of experimental constraints and theoretical consistency
conditions: perturbativity, unitarity and vacuum stability, as well
as from the EW precision data, the low-energy flavor-physics inputs
and the Higgs mass regions ruled out by the LEP and Tevatron direct
searches -- cf. e.g. Ref.
~\cite{constraints_general,superiso,unitarity,vacuum}.


\section{One-loop Higgs boson production in $ \APelectron\Pelectron$}

In the following we present a full-fledged one-loop analysis of the
CP-conserving pair production of neutral Higgs bosons
($\APelectron\Pelectron \to 2h=\hzero\Azero/\Hzero \Azero$),
alongside with the associated Higgs/$\PZ^0$ boson
(\emph{Higgs-strahlung}) channels within the 2HDM. On top of the
complete set of $\mathcal{O}(\alpha^3_{ew})$ corrections, we also
retain the leading $\mathcal{O}(\alpha^4_{ew})$ terms which stem
from the (squared of the) $\mathcal{O}(\lambda^2_{3h})$
Higgs-mediated contributions. Renormalization of the SM fields and
coupling constants is performed in the conventional on-shell scheme
in the Feynman gauge
 \cite{renorm_sm}. A dedicated extension of the
on-shell scheme is worked out for the 2HDM Higgs sector
in\,\cite{loop1}.
Phenomenological constraints are implemented by interfacing our
numerical codes with the packages \textit{2HDMCalc} \cite{2hdmcalc},
\textit{SuperISO} \cite{superiso} and \textit{HiggsBounds}
\cite{higgsbounds}, altogether with several complementary in-house
routines. As for the algebraic calculation and numerical evaluation
of the cross-sections under study, we have employed the standard
computational software \textit{FeynArts}, \textit{FormCalc} and
\textit{LoopTools} \cite{feynarts}. Representative choices for the
Higgs boson spectrum are sorted out in two sets as indicated in
Table \ref{tab:mass2hdm}.

\begin{figure}[t]
 \begin{center}
   \includegraphics[scale=0.5]{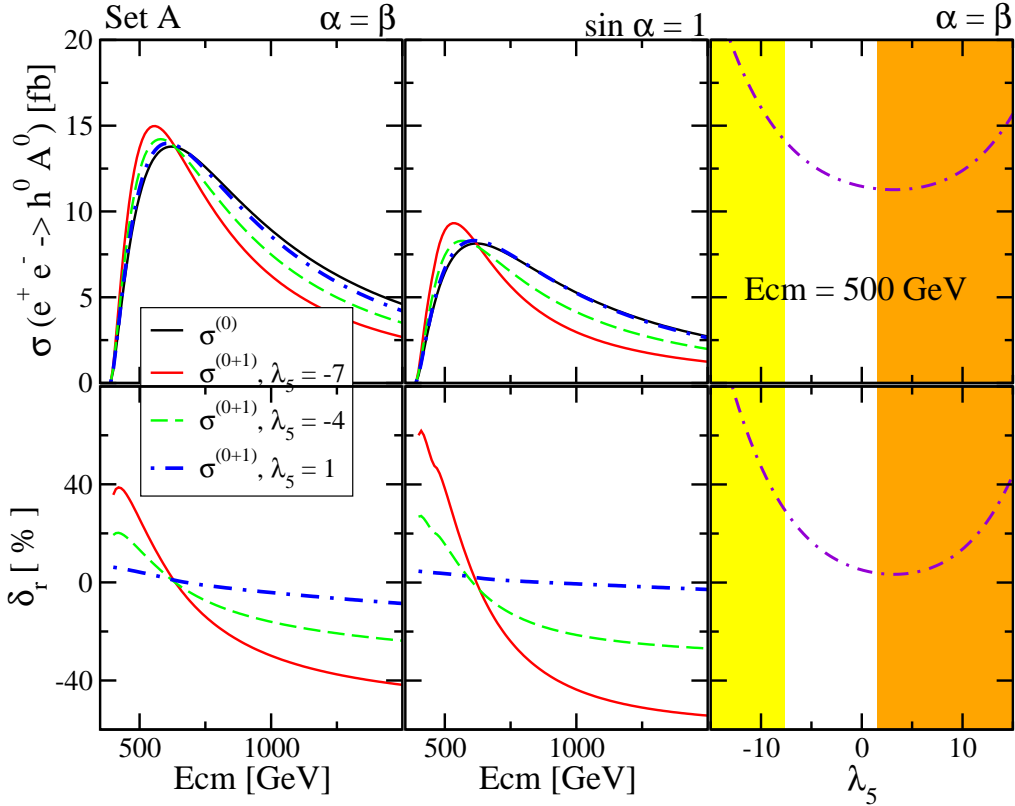}
\caption{Total cross-section $\sigma(\APelectron\Pelectron
\to\hzero\Azero)$ (in fb) and relative one-loop correction
$\delta_r$ (in $\%$) for Set A of Higgs boson masses. Left and
central panels display these quantities as a function of $\sqrt{s}$
(indicated as Ecm). The results are shown within three different
values of $\lambda_5$ at fixed $\tan\beta = 1.2$ (compatible with
the lower $\tan\beta$ bound from $B_d^0 - \bar{B}_d^0$
data\,\cite{superiso}) and for the representative choices
$\alpha=\beta$ (maximum tree-level coupling) and $\alpha=\pi/2$
(fermiophobic limit of $\hzero$ for type-I 2HDM). In the right panel
we show their evolution in terms of $\lambda_5$ at $\sqrt{s}=500$
GeV. The shaded areas on the left (resp. right) are excluded by
unitarity (resp. vacuum stability).}
\end{center}
\label{fig:2h1}
\end{figure}

\begin{table}[thb]
\begin{center}
\begin{tabular}{|c||c|c|c|c|}
\hline
        2HDM & $M_{\hzero}$ (GeV)& $M_{\Hzero}$ (GeV)& $M_{\Azero}$ (GeV)& $M_{\Hpm}$(GeV)  \\
\hline\hline
Set A & $130$ &  $200$ &  $260$  & $300$  \\
Set B & $115$ &  $165$ &  $100$  & $105$  \\
\hline
\end{tabular}
\end{center}
\caption{\footnotesize{Sets of 2HDM Higgs boson masses used
throughout the calculation. Owing to the flavor constraints on
$M_{\Hpm}$\,\,\cite{superiso}, Set B is only possible for type-I
2HDM's, whereas Set A is possible for both type-I and type-II. The
mass sets satisfy the custodial symmetry bound
$|\delta\rho|<10^{-3}$ -- cf. Ref.\cite{loop1}.}}
\label{tab:mass2hdm}
\end{table}

Our interest here is basically threefold, namely: i) to seek for
regions within the 2HDM parameter space sourcing large quantum
corrections, which we shall quantify through the ratio $\delta_r
\equiv \sigma^{(1)}/\sigma^{(0)}$, where
$\sigma^{(1)}=\sigma-\sigma^{(0)}$ is the one-loop correction with
respect to the tree-level value; ii) to evaluate their impact on the
overall 2h and hZ production rates (cf. Figs. 1-3); and iii) to
correlate these effects to the strength of the $3h$ self-couplings.

Worth noticing is that the Higgs/gauge boson couplings ($hZZ,hAZ$)
driving these processes at the leading-order are anchored by the
gauge symmetry, and hence take the same form in both the 2HDM and
the MSSM. Genuine differences between both models should thus be
probed through the study of quantum effects -- among which the
enhanced $3h$ self-interactions of the 2HDM could rubber-stamp a
very distinctive imprint.

\begin{figure}[t]
 \begin{center}
   \includegraphics[scale=1.1]{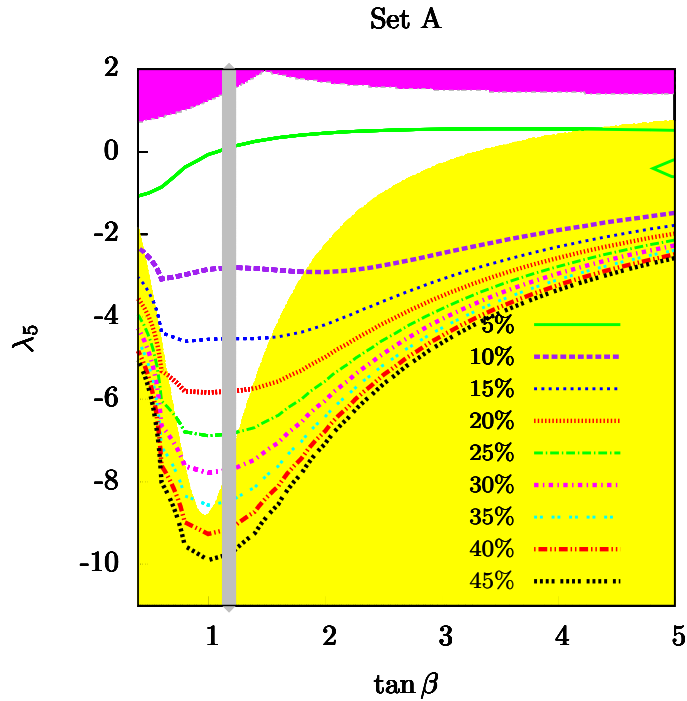}
\caption {Radiative corrections $\delta_r$ (\%) to the total
cross-section $\sigma(\APelectron\Pelectron \to\hzero\Azero)$ as a
function of $\tan\beta$ and $\lambda_5$, for Set A of Higgs boson
masses, $\alpha=\beta$ and $\sqrt{s} = 0.5\, \TeV$. The shaded areas
in the top (resp. bottom) account for the vacuum
stability\,\cite{vacuum} (resp. unitarity\,\cite{unitarity})
conditions, while the vertical grey band depicts the lower limit
$\tan\beta \simeq 1.18$ ensuing from $B_d^0 -
\bar{B}_d^0$\,\cite{superiso}. }
\end{center}
\label{fig:2h2}
\end{figure}
Let us concentrate on Set A of Higgs boson masses for the analysis
of this section. This set is possible in both type-I and type-II
2HDM's\,\cite{2hdmrev}. (Set B, characterized by lighter Higgs
masses, will be used for $\gamma\gamma$ physics in the next section;
it is only allowed for type-I models.) The cross-section for the
particular channel $\hzero\Azero$ as a function of the center of
mass energy $\sqrt{s}$ (Ecm), displayed for a few values of
$\lambda_5$, and also as a function of $\lambda_5$ at fixed
$\sqrt{s}$, is shown in Fig.\,1,
together with the relative quantum correction $\delta_r$. The
rightmost panel of this figure illustrates the expected $\sigma \sim
\sigma_0+{\cal O}(\lambda_5^2)+{\cal O}(\lambda_5)$ behavior
triggered by the triple Higgs boson self-interactions\,\cite{loop1}.
The cross-sections at one-loop lie in the approximate range of
$2-15\,\femtobarn$ for $\sqrt{s} = 0.5\,\TeV$ -- rendering
$10^3-10^4$ events per $500\,\invfb$ of integrated luminosity. The
corresponding quantum effects are large and positive for $\sqrt{s}$
around the nominal startup energy of the ILC, i.e. $0.5\,\TeV$, but
become rapidly negative for $\sqrt{s} \gtrsim 0.6\,\TeV$ and stay
highly so in the entire ${\cal O}(1)$ TeV regime. This sign flip in
combination with the behavior of the quantum effects on the
Higgs-strahlung processes (see below) gives an important
experimental handle on the physics of Higgs production in the 2HDM.
Although we have used Set A for the analysis, the behavior of
$\delta_r$ turns out to be fairly independent of the details of the
Higgs mass spectrum, the particular type of 2HDM and the specific
channel under analysis ($\hzero\Azero$ or $\Hzero\Azero$).

A dedicated study of the quantum effects $\delta_r$  as a function
of $\tan\beta$ and $\lambda_5$ is explored in
Fig.\,2, at fixed $\sqrt{s}=500$\,GeV. We can also appreciate in it
the interplay with the unitarity bounds (lower area, in yellow) and
the vacuum stability conditions (upper area, in purple). Notice that
the former disallows simultaneously large values of $\tan\beta$ and
$\lambda_5$, whereas the latter enforces $\lambda_5 \lesssim 1$, but
mainly in the negative range: $-10<\lambda_5<0$. The largest
attainable quantum effects ( $|\delta_r|\sim 20-60\,\%$) are
localized in a valley-shaped region centered at $\tan\beta\gtrsim 1$
deep in the allowed $\lambda_5<0$ range (cf. Fig.\,2). Here a subset
of $3h$ self-couplings becomes substantially augmented -- their
strength growing with $\sim |\lambda_5|$ -- and stands as a
preeminent source of radiative corrections via Higgs-boson mediated
one-loop corrections to the $h\Azero\PZ^0$ vertex. As a result this
interaction vertex, which is purely gauge at the tree-level, can be
drastically modified at the one-loop order. There is, however, the
rigid lower bound  $\tan\beta\gtrsim 1$ from $B_d^0 - \bar{B}_d^0$
oscillations\,\cite{superiso} (see the vertical grey band in
Fig.\,2), which further restricts the valley-shaped region and
finally leaves less than half of its original allowance for the
largest possible quantum effects.
\begin{figure}[t!]
 \begin{center}
   \includegraphics[scale=0.5]{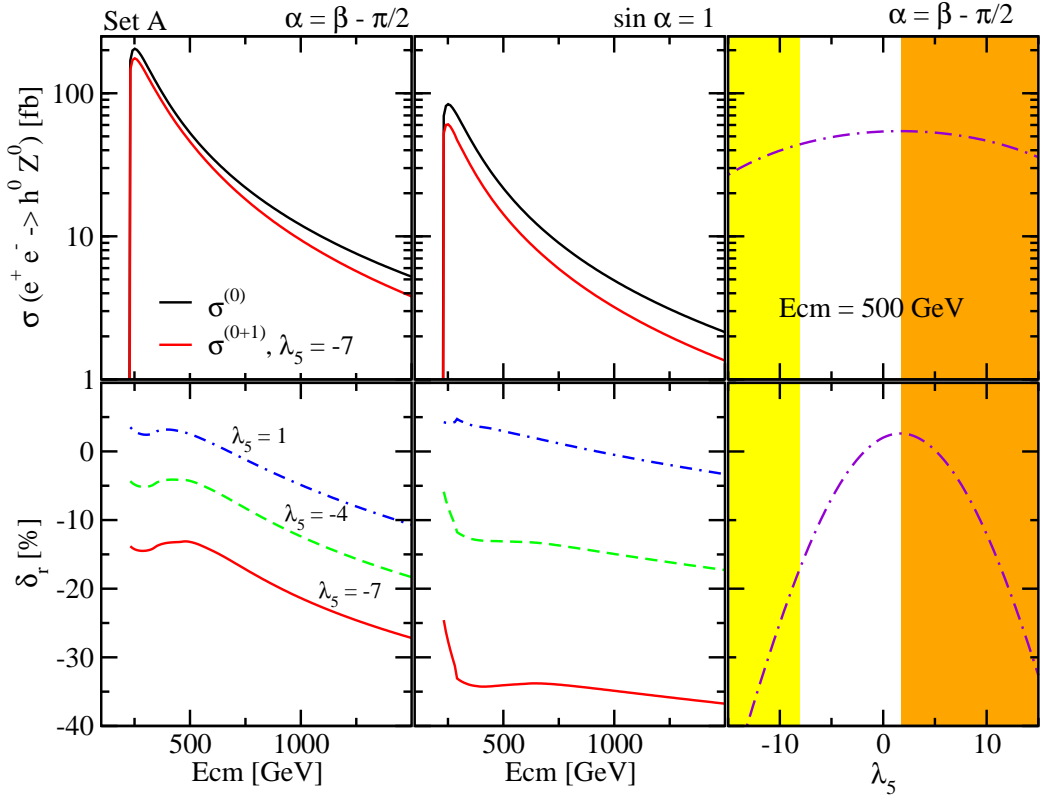}
\caption{Total cross-section $\sigma(\APelectron\Pelectron \to\hzero
Z^0)$ (in fb) and relative one-loop correction $\delta_r$ (in $\%$)
for Set A of Higgs boson masses. Left and central panels display
these quantities as a function of $\sqrt{s}$, for three different
values of $\lambda_5$, $\tan\beta = 1.2$ and for the representative
choices $\alpha=\beta-\pi/2$ and $\alpha=\pi/2$. In the right panel
we show their evolution in terms of $\lambda_5$ at $\sqrt{s}=500$
GeV. The shaded areas on the left (resp. right) are excluded by
unitarity (resp. vacuum stability).}
\end{center}
\label{fig:hz}
\end{figure}

In addition to the pairwise production of Higgs bosons we find the
more traditional Higgs-strahlung channels ($\APelectron\Pelectron
\to \hzero Z^0, \Hzero Z^0$)\,\cite{loop2}, i.e. the 2HDM analog(s)
of the so-called Bjorken process in the SM\,\cite{Bjorken76}.
As reported in Fig.\,3 (left and central panels), we obtain typical
cross sections in the ballpark of $\sigma(\hzero Z^0) \sim
\mathcal{O}(10-100) \,$fb, which may undergo substantial (and mostly
negative) radiative corrections up to order $\delta_r \sim -50\%$
for large $3h$ self-coupling enhancements -- these being preferably
realized for $\tan\beta = \mathcal{O}(1)$ and $|\lambda_5| \sim
\mathcal{O}(10)$. The trademark negative sign of the leading quantum
effects can be tracked down to the dominance of the finite
wave-function corrections to the external Higgs boson fields, this
being the only contribution at one loop which retains a quadratic
dependence on $\lambda_{3h}$. The rightmost panel of Fig.\,3 nicely
illustrates the characteristic $\sigma \sim \sigma_0-{\cal
O}(\lambda_5^2)+{\cal O}(\lambda_5)$ behavior induced by the triple
Higgs boson self-interactions, which in this case produce dominant
(and negative) quantum effects of ${\cal O}(\lambda_5^2)$ from the
wave function renormalization as long as the regime $|\lambda_5|>1$
is well attained\,\cite{loop2}. As hinted before, the correlation of
large negative quantum effects on the Higgs-strahlung channels with
the presence of significant positive (for $\sqrt{s}\lesssim 500$
GeV) or negative (for $\sqrt{s}>600$ GeV) quantum effects on the
double Higgs production channels (cf. Fig. 1) could eventually lead
to a robust quantum signature of (non-supersymmetric) 2HDM physics
in a Linear Collider.

\section{One-loop Higgs boson production from $ \gamma\gamma$}

Direct $\gamma\gamma$ collisions may be realized through Compton
backscattering of high energetic laser pulses off the original
$e^+e^-$ linac beams\, \cite{gammacoll}. This alternative running
mode opens up a plethora of complementary experimental strategies
for a Linear Collider. In particular, it may provide a pristine
insight into the loop-induced $\gamma\gamma h$ coupling -- and so to
the underlying structure of the Higgs sector. This effective
interaction ensues from an interplay of gauge boson, fermion and
charged scalar one-loop contributions \cite{ellis}. In the 2HDM the
charged Higgs-mediated effects, which are directly sensitive to the
$3h$ self-coupling $\lambda_{h H^+H^-}$, along with the modified
Higgs/fermion and Higgs/gauge boson interactions, are responsible
for a highly characteristic phenomenological pattern. To illustrate
it, we shall concentrate on the following quantities: i) the total
(unpolarized and averaged) cross-section
$\langle\sigma_{\gamma\gamma\to h}\rangle(s)$, which results from
the convolution of the ``hard'' scattering cross-section
$\hat{\sigma}(\gamma\gamma \to h)$  with the photon luminosity
distribution (describing the  effective $e^{\pm}\to\gamma$
conversion of the primary linac beam). For the latter we use the
parametrization included in the standard package
CompAZ~\cite{compaz}); ii) the relative strength $r$ of the
effective $\Pphoton\Pphoton h$ interaction normalized to the SM
(upon identifying $M_{\PHiggs_{SM}}$ with $M_{\hzero}$), namely $r
\equiv g_{\Pphoton\Pphoton h}/g_{\Pphoton\Pphoton H}^{\rm SM}$, by
which we aim at quantifying the departure of the genuine dynamical
features of the 2HDM with respect to the SM.

\begin{figure}[t]
 \begin{center}
\hspace{-0.4cm}   \includegraphics[scale=1.3]{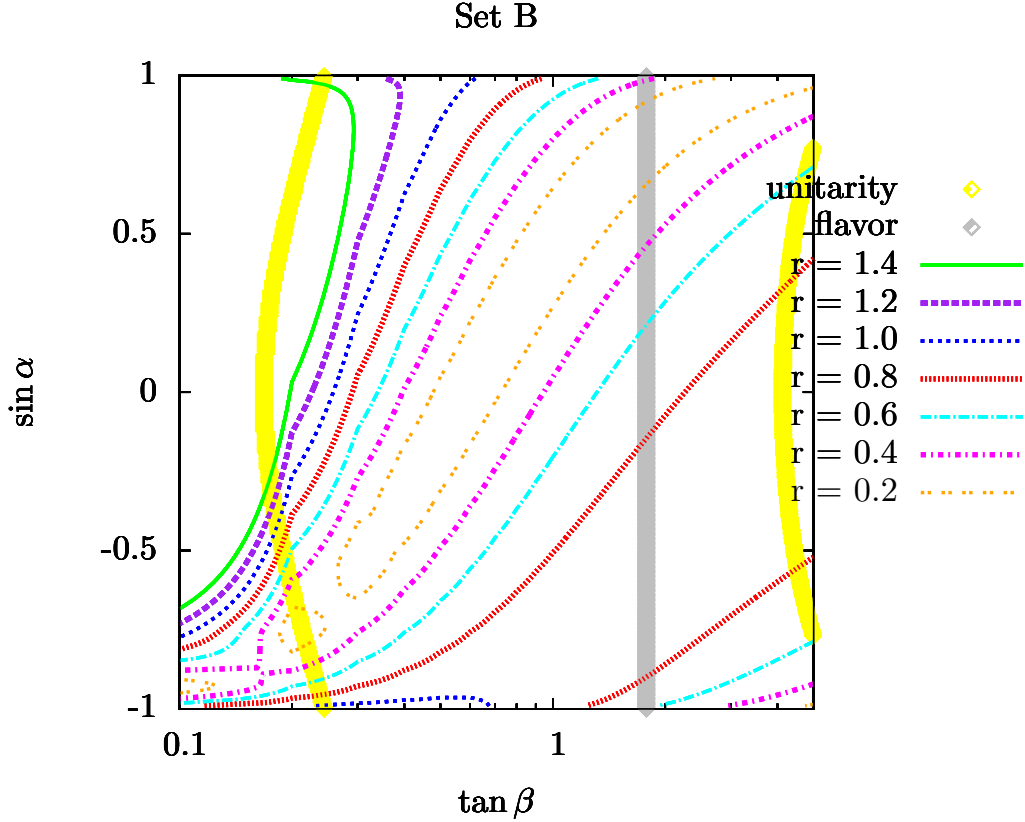}
 \end{center}
\caption{Effective $\Pphoton\Pphoton\hzero$ coupling strength in the
2HDM normalized to the SM, $r \equiv g_{\Pphoton\Pphoton
\hzero}/g_{\Pphoton\Pphoton H}^{\rm SM}$, in terms of $\sin\alpha$
and $\tan\beta$. The yellow strips signal the lower and upper bounds
stemming from unitarity, while the grey vertical band ensues from
the $\bar{B}_d^0-B_d^0$ constraints at the $3\sigma$ level.}
\label{fig:2photon}
\end{figure}

\begin{figure}[t]
 \begin{center}
\includegraphics[scale=0.6]{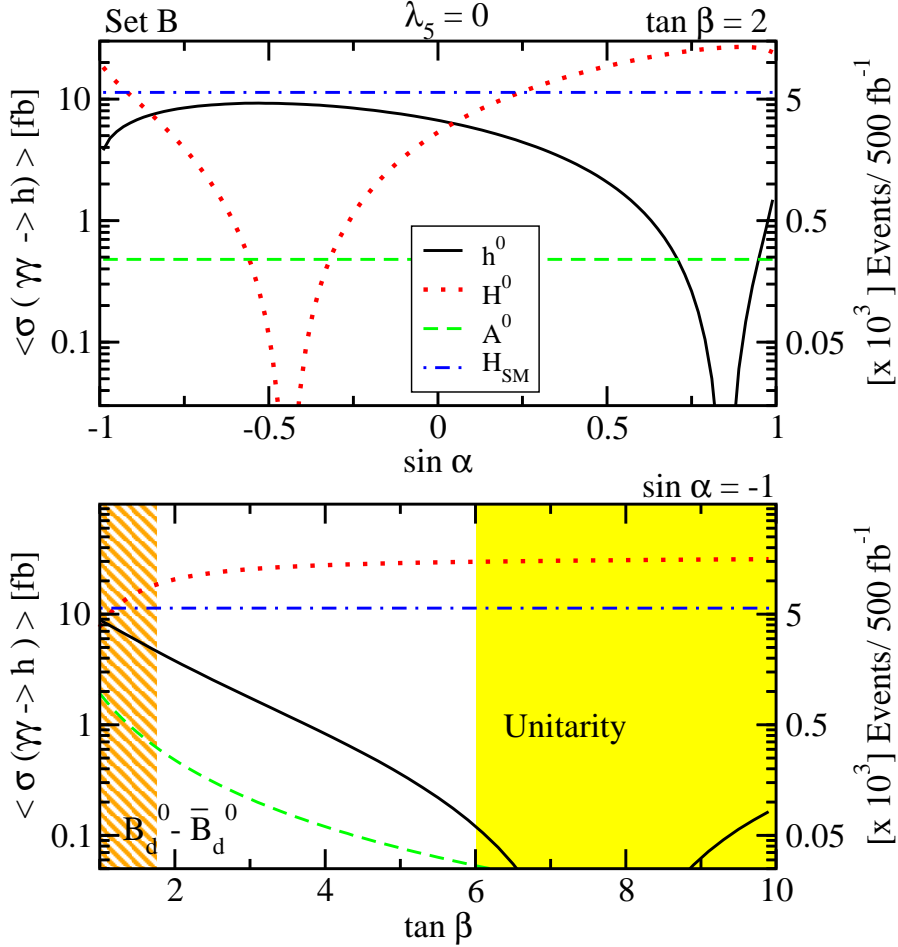}
 \end{center}
\caption{Averaged cross-section $\langle\sigma_{\gamma\gamma\to
h}\rangle(\sqrt{s} = 500\,GeV )$ (in fb) as a function of
$\tan\beta$ (top panel) and $\sin\alpha$ (bottom panel), for Set B
of Higgs boson masses and $\lambda_5 = 0$. The yellow-shaded (resp.
orange-dashed) areas are disallowed by unitarity (resp.
$\bar{B}_d^0-B_d^0$) constraints.} \label{fig:sigmatanbeta}
\end{figure}

%
%

The results of our numerical analysis of $r$ are displayed in
Fig.~\ref{fig:2photon}. A first observation is that, unsurprisingly,
the profile of $g_{\gamma\gamma \hzero}$ is mainly modulated by the
strength of the $3h$ coupling $\lambda_{\hzero\PHiggs^+\PHiggs^-}$.
This is a consequence of a destructive interference between the
contributions from the loop diagrams triggered by the Higgs boson
self-interaction $\hzero\PHiggs^+\PHiggs^-$ and those induced on the
one hand by the gauge bosons and those driven by the Yukawa
interactions with neutral Higgs bosons and fermions on the other
(cf. Fig. 2 of \cite{photon}). The impact of such interference is
well visible for wide areas of the parameter space where the strong
departures from $r \simeq 1$ can be traded to suppressions of order
$40-60\%$ for the effective $g_{\gamma\gamma h}$ interaction as
compared to the SM. Away from these depleted domains, maximum
cross-sections of order $\sigma \sim \mathcal{O}(10)$ fb -- viz. up
to a few thousand events per $500\invfb$ -- are attainable for both
$\hzero$ and $\Hzero$ (cf. Fig.~\ref{fig:sigmatanbeta} for the
dependence of the cross section with $\sin\alpha$ and $\tan\beta$).
Interestingly enough, the optimal production rates are nicely
complementary for the \CP-even channels $\gamma\gamma\to\hzero$ and
$\gamma\gamma\to\Hzero$, as a result of the inverse correlation of
the respective $\lambda_{h\PHiggs^+\PHiggs^-}$ self-couplings (see
Table II of \cite{loop1}) -- and hence of the dominant interference
effects. In contrast, owing to the CP-odd nature of $\Azero$, the
$\gamma\gamma\Azero$ channel appears to be rather featureless and
has a milder numerical impact.

Besides the $3h$ self-couplings, further 2HDM mechanisms could
contribute, at least in principle, to shift the $r$ ratio from its
canonical value $r = 1$. For example, $r > 1$ could be achieved for
$\tan\beta \sim 0.2 - 0.3$ (cf. Fig.~\ref{fig:2photon}) as a result
of the enhanced Higgs-top ($\sim 1/\tan\beta$) Yukawa coupling. In
practice, however, the ${B}_d^0-\bar{B}_d^0$ bounds exclude this
possibility. Significant departures from the SM may arise as well
within large $\lambda_5$ scenarios, such as those analyzed in
\cite{photon}. This possibility is nonetheless highly disfavored if
the unitarity conditions are included in their most restrictive
version \cite{unitarity}.

All in all, trademark hints of a 2HDM structure may emerge from
$\gamma\gamma \to h$, mainly through a missing number of events with
respect to the SM predictions -- as long as the overall rates are
still large enough to be efficiently discriminated from the dominant
background process, $\gamma\gamma \to b\bar{b}$. We have shown that
this situation can still be realized in sizeable regions of the
parameter space, provided the $3h$ self-couplings take on moderate
values (viz. $\lambda_{3h} ~\sim \mathcal{O}(10^2)$ GeV) preserving
unitarity. In contrast, in the MSSM the genuine supersymmetric
(slepton/squark-mediated) contributions to $g_{\gamma\gamma h}$ can
only induce rather tempered quantum effects as compared to the
general 2HDM, the main reason being the absence of potentially large
$3h$ self-couplings, and hence the lack of a mechanism able to
prompt the characteristic interference pattern that we have
identified above. Although alternative enhancing effects on
$\gamma\gamma\to h$ within the MSSM are
possible\,\cite{previousmssm}, e.g. through (light) stop-mediated
loops with large trilinear $(A_t)$ couplings and sizable mass
splittings between their chiral components $\tilde{t}_1, \tilde{t}_2
$, they are nevertheless comparably weaker. In fact, these effects
are always pulled down by inverse powers of the SUSY breaking scale
and are further limited by the stringent limits on the squark and
Higgs boson masses, as well as from $B$-meson physics -- cf. the
recent Ref.~\cite{photonmssm} for a fully updated MSSM analysis and
a comparison with the general 2HDM case. The bottom-line is that
mild deviations from $r=1$ of order $-5\%$ characterize the typical
MSSM scenarios. In this sense, it is worth emphasizing that, even if
a pattern of the sort $r \lesssim 1$ would overlap with the 2HDM
predictions for small $3h$ self-couplings (viz. $\lambda_{3h} \sim
10$ GeV), the correlation of $\gamma\gamma \to \hzero$ and
$\gamma\gamma \to \Hzero$ could still help to disentangle both
models. Indeed, in the 2HDM both \CP-even channels could
simultaneously yield $\sigma \sim 1-10$ fb (see Fig.\,5), whilst
such situation is definitely precluded in the MSSM owing to the SUSY
restrictions on the Higgs boson mass splittings --  see
~\cite{photonmssm} for details.

\section{Discussion and Conclusions}

In this work, we have described several phenomenogical aspects of
one-loop Higgs boson production processes within the general 2HDM at
the future Linear Colliders. We have revisited our previous results
and cast them into a common, fully updated framework, including the
most recent set of theoretical and experimental constraints
presently available in the literature -- most significantly those
stemming from unitarity and flavor physics. Our brought-to-date
analysis keeps on highlighting the truly instrumental role reserved
to the future linac facilities, and also the great degree of
complementarity of the $\APelectron\Pelectron$ and
$\Pphoton\Pphoton$ running modes.

We have provided detailed, quantum-corrected predictions for the
exclusive pairwise production of Higgs bosons $\APelectron\Pelectron
\to 2h$ as well as for the Higgs-strahlung channels
$\APelectron\Pelectron \to hZ$ at $\mathcal{O}(\alpha^3_{ew})$ and
leading $\mathcal{O}(\alpha^4_{ew})$.
In the case of $2h$ production, we have shown that the radiative
corrections can reach the level of $|\delta_r| \sim 50\%$ for
enhanced $3h$ self-couplings -- mostly for low $\tan\beta$ and
$|\lambda_5| \sim \mathcal{O}(10)$ -- and hence give rise to a
substantial positive boost with respect to the tree-level
expectations (around $\sqrt{s} \simeq 500$ GeV) or suppression (for
$\sqrt{s}> 600$ GeV). In the case of the hZ final states the quantum
effects can be of the same order of magnitude, but they are negative
for essentially the whole $\sqrt{s}$ range. Optimal rates for these
processes lie in the ballpark of ${\cal O}(10-100)$ fb and can be
attained for typical Higgs masses in the $100 -300 \,\GeV$ range.

No less crucial is the role of the $3h$ self-couplings in the
production of a single Higgs boson from direct $\gamma\gamma$
collisions. A trademark suppression of $\sigma(\gamma\gamma \to h)$
with respect to the SM predictions is singled out, and its origin
traced back to the interference effects between the different
one-loop contributions, these being critically modulated by the
strength of the $3h$ self-couplings $h\PHiggs^+\PHiggs^-$. In
regions of the parameter space for which this depletion is moderate
(viz. $\lambda_{h\PHiggs^+\PHiggs^-} \sim \mathcal{O}(10-100)$ GeV),
still significant cross-sections up to $\sigma \sim 10$ fb may be
retrieved for both $\hzero$ and $\Hzero$. From the experimental
point of view, the opportunities for accessing this kind of final
states are deemed to be excellent. The decay signatures of the
neutral CP-even states would essentially boil down to $\hzero \to
b\bar{b}/\tau^+\tau^-$ or $\hzero \to VV \to 4l, 2l +
\slashed{E}_T$, all of them allowing for a comfortable tagging in
the clean linac environment. Interestingly enough, the described
phenomenology is particularly distinctive of a non-supersymmetric
2HDM structure. In the MSSM, potential enhancements cannot be
triggered by the Higgs self-interactions -- which are anchored by
the gauge couplings -- but instead by the Yukawa interactions of the
Higgs bosons and the sfermions. These give rise, in general, to
rather tempered quantum effects, as compared to the sizable
corrections that are spotlighted for the 2HDM. Likewise, the
conditions that SUSY dictates on the Higgs spectrum may be of
relevance here. For instance, genuine indication of non-standard,
non-SUSY Higgs physics may come from the simultaneous observation of
$\gamma\gamma \to \hzero$ and $\gamma\gamma \to \Hzero$; both
channels may yield $\mathcal{O}(10^3)$ events per $500\invfb$ -- a
situation which could never be ascribed to the MSSM, as the mass
splitting between the two Higgs bosons is enforced to be much
larger. Finally, the combined analysis of these signatures together
with complementary multi-Higgs production channels (cf.
Ref.~\cite{loop1}) could unveil a characteristic pattern of
signatures for different values of $\sqrt{s}$. If confirmed, it
would point to a non-standard, non-supersymmetric origin.
Conversely, if the two \CP-even states would be produced at
measurable rates differing by, say, one order of magnitude, this
could be compatible with MSSM Higgs physics (see \cite{photonmssm}
for details), but it would require a detailed dijet invariant mass
reconstruction to resolve the spectrum and check if it is compatible
with the MSSM constraints.

\vspace{0.25cm}
 \noindent
\textbf{Acknowledgments}\,\, JS would like to thank the organizers
of the LC 2010 workshop at the INFN-Laboratori Nazionali di Frascati
for the kind invitation to present this review. This work has been
supported in part by DIUE/CUR Generalitat de Catalunya under project
2009SGR502; by MEC and FEDER under project FPA2010-20807, and by the
Spanish Consolider-Ingenio 2010 program CPAN CSD2007-00042.


\newcommand{\JHEP}[3]{ {JHEP} {#1} (#2)  {#3}}
\newcommand{\NPB}[3]{{\sl Nucl. Phys. } {\bf B#1} (#2)  {#3}}
\newcommand{\NPPS}[3]{{\sl Nucl. Phys. Proc. Supp. } {\bf #1} (#2)  {#3}}
\newcommand{\PRD}[3]{{\sl Phys. Rev. } {\bf D#1} (#2)   {#3}}
\newcommand{\PLB}[3]{{\sl Phys. Lett. } {\bf B#1} (#2)  {#3}}
\newcommand{\EPJ}[3]{{\sl Eur. Phys. J } {\bf C#1} (#2)  {#3}}
\newcommand{\PR}[3]{{\sl Phys. Rept. } {\bf #1} (#2)  {#3}}
\newcommand{\RMP}[3]{{\sl Rev. Mod. Phys. } {\bf #1} (#2)  {#3}}
\newcommand{\IJMP}[3]{{\sl Int. J. of Mod. Phys. } {\bf #1} (#2)  {#3}}
\newcommand{\PRL}[3]{{\sl Phys. Rev. Lett. } {\bf #1} (#2) {#3}}
\newcommand{\ZFP}[3]{{\sl Zeitsch. f. Physik } {\bf C#1} (#2)  {#3}}
\newcommand{\MPLA}[3]{{\sl Mod. Phys. Lett. } {\bf A#1} (#2) {#3}}
\newcommand{\JPG}[3]{{\sl J. Phys.} {\bf G#1} (#2)  {#3}}
\newcommand{\JPCF}[3]{{\sl J. Phys. Conf. Ser.} {\bf G#1} (#2)  {#3}}
\newcommand{\FDP}[3]{{\sl Fortsch. Phys.} {\bf G#1} (#2)  {#3}}
\newcommand{\CPC}[3]{{\sl Com. Phys. Comm.} {\bf G#1} (#2)  {#3}}

\end{document}